\documentstyle[preprint,prl,aps]{revtex}         
\def\noi{\noindent}
\def\pc{\,{\rm pc}}
\def\kpc{\,{\rm kpc}}
\def\gev{\,{\rm GeV}}
\def\cm{\,{\rm cm}}
\def\km{\,{\rm km}}
\def\s{\,{\rm s}}

\def\kmps{\km\s^{-1}}

\def\gevpcc{\gev\cm^{-3}}
\def\msun{\,{\rm M}_\odot}
\def\msunppcsq{\msun\pc^{-2}}
\def\msunppccube{\msun\pc^{-3}}
\def\dm{_{\rm DM}}

\def\rmsveldm{\langle v^2\rangle_{\dm}^{1/2}}
\def\rmsveldmsq{\langle v^2\rangle_{\dm}}

\begin{document}

\draft

\title{ The Dispersion-Velocity of Galactic Dark Matter Particles }

\author{ 
R.Cowsik$^{1,2,3,}$\cite{ast},
Charu Ratnam$^{1,4,}$\cite{dag},
and 
P.Bhattacharjee$^{1,}$\cite{ddag}
}

\address{ $^1$ Indian Institute of Astrophysics, 
 Koramangala, Bangalore 560 034. India.}

\address{ $^2$ McDonnell Center for Space Sciences, 
 Washington University, St Louis, MO 63130. USA.}

\address{ $^3$ Tata Institute of Fundamental Research, 
 Homi Bhabha Road, Bombay 400 005. India.}

\address{ $^4$ Joint Astronomy Program, Indian Institute of Science, 
 Bangalore 560 012. India.}

\maketitle

\begin{abstract}

The self-consistent spatial distribution of particles of
Galactic dark matter is derived including their own gravitational
potential, as also of that of the visible matter of the Galaxy. In order
to reproduce the observed rotation curve of the Galaxy the value of 
the dispersion velocity of the dark matter particles, $\rmsveldm$,  
should be $\sim 600\kmps$ or larger. 

\end{abstract}

\tighten

More than 20 years ago, it was suggested \cite{cm} that weakly
interacting particles of non-zero rest mass which decouple from 
radiation and matter early after the Big Bang would form an invisible 
gravitating background of dark matter (DM) 
around galactic systems. Even though at that time the only
available candidates for these particles were the neutrinos of the muon and 
electron flavors, the idea itself became the paradigm under which the
newly discovered particles like the tau-neutrino and newly hypothesised  
particles within the context of possible physics 
beyond the Standard Model of particle physics could be incorporated. 
Also, during the latter half of the intervening twenty
years, we have witnessed a tremendous growth in the experimental effort
towards direct detection of these particles in the laboratory. The
experiments are aimed at observing the effects of the impact of mainly the
more massive candidate particles of DM with targets maintained at 
cryogenic temperatures
which facilitate the observation of the tiny amount of energy deposited
in the process against the background 
generated by internal and external radioactivity and by the cosmic
rays. These developments are reviewed in detail by Trimble \cite{vt}, 
Primack, Seckel and Sadoulet \cite{pss}, Caldwell
\cite{doc}, and Price \cite{pb} . 

The interpretation of these experiments to derive constraints                 
on the properties of the unknown particles constituting a halo of dark
matter in and around the Galaxy requires assumptions about the density 
and spectrum of velocities of the DM particles in the solar neighbourhood. 
These parameters have 
been obtained thus far by describing the DM halo as a single component 
isothermal sphere which is truncated at a particular radius \cite{raf}.
The normalization for the density of DM particles comes from an analysis
originally suggested by Oort \cite{jho} in which the observed spatial- and
velocity distribution of stars near the solar system indicate a DM density 
of $\sim 0.3 \gevpcc$ in the solar neighbourhood; Bahcall \cite{jnb} 
gives a detailed account of this procedure. The 3-dimensional dispersion 
velocity of the DM particles, $\rmsveldm$, has not been determined, however. 
It is customary to take recourse to the virial result pertaining to an 
isotropic isothermal sphere \cite{bt} and set $\rmsveldm=\sqrt{3\over2}\,
\Theta_\infty$, where $\Theta_\infty$ is the asymptotic value of the circular 
rotation speed. Since $\Theta_\infty$ for the Galaxy is not known, the 
usual practice is to {\it assume} that the rotation curve of the 
Galaxy \cite{bg,fbs,ft}, $\Theta(R)$, is flat from
$R\sim5\kpc$ out to $R\gg R_0 
\approx 8.5\kpc$ (here and below $R$ denotes 
the galactocentric distance in the plane of the Galaxy, $R_0$ being the 
sun's position), and set $\Theta_\infty\approx\Theta (R_0)\approx 220\kmps$, 
the rotation speed near the solar system. This yields 
$\rmsveldm\approx 270\kmps$, 
which is the value usually assumed in most studies of 
issues related to Galactic DM. However, as noted in the recent review by 
Fich and Tremaine \cite{ft}, ``Much of the data indicates that the rotation 
curve continues to rise beyond $R_0$''. Thus the estimate $\rmsveldm\sim 
270\kmps$ derived by assuming $\Theta_\infty=\Theta (R_0)$ is 
uncertain. Moreover, the assumption of a pure isothermal sphere for 
the description of the dark matter halo neglects the possible deviation 
from spherical symmetry induced by the disk-like distribution of the 
visible matter. 

Keeping these points in mind, we focus attention on the observed 
rotation curve of the Galaxy, and develop a theoretical framework, 
the salient features of which are: (a) A model for the Galaxy comprising 
of visible matter and particles of DM with a {\it self-consistent} 
inclusion of their gravitational interactions, and (b) Departure from 
spherical symmetry due to the disk-like distribution of the visible 
matter which will be treated as axially symmetric. The quantity $\rmsveldm$ 
appears as a free parameter in our framework and is determined by comparing 
the theoretical rotation curve with the observed data.   

We adopt well-established models to describe the density distribution of 
the normal visible matter and the resulting gravitational potential. In 
this {\it Letter} we present our results for a two-component model of the 
visible matter consisting of a spheroidal bulge
\cite{bt,co,kg} with density 
$\rho_s(r)$, and an axisymmetric disk \cite{kg} with 
density $\rho_d(R,z)$: 
\begin{eqnarray}
\rho_s(r)&=&{\rho_0\over\left(1+{r^2\over
a^2}\right)^{3/2}} \\
\rho_d (R,z)&=&{\Sigma_0\over2h}e^{-(R-R_0)/R_d}\,e^{-\vert
z\vert/h}  
\end{eqnarray}
where $r=\left(R^2+z^2\right)^{1/2}$, and 
$\Sigma_0\equiv\int_{-\infty}^{\infty}\rho_d(R_0,z)dz$ is the 
disk surface density at the solar position, $z$ being the vertical distance    
from the plane of the disk. The values of the parameters are given by 
\cite{co,kg} $a=0.103\kpc$, $R_d=3.5\kpc$, $h=0.3\kpc$, and 
$\rho_s(R_0)=7\times10^{-4}\msunppccube$. 
(Note that the rotation curve in the 
outer regions of the Galaxy is relatively insensitive to the spheroid 
parameters). 
There are conflicting reports 
on the value of $\Sigma_0$: Whereas Kuijken and Gilmore
\cite{kg} (KG) suggest 
$\Sigma_0\sim40\msunppcsq$ on the basis of data on $\sim$ 512 K-dwarf 
stars, Bahcall et al \cite{bss,bfg} in their reanalysis of
essentially the same 
data suggest a number for $\Sigma_0$ which is about twice as large. In 
our calculations we consider values of $\Sigma_0$ in the range 
(40--80)$\msunppcsq$. The estimate of the local surface density 
of the Galactic disk due to the identified matter such as visible stars 
is $\sim 48\pm 8\msunppcsq$. Thus Bahcall et al's kinematical estimate of   
$\Sigma_0$ seems to indicate the presence of a substantial amount of unseen 
matter in the Galactic disk, whereas KG's estimate is consistent with no 
disk dark matter. (Note that analyses of Refs.\cite{kg,bss,bfg} are all 
based on 
1-dimensional solutions to the Boltzman equation, which, in the given 
situation, are strictly valid for an infinite disk only). In any case, the  
dark matter associated with the disk is likely to be dissipational in 
contrast to that constituting the extended halo which would be collisionless 
and non-dissipative. We are concerned with this latter type of dark matter 
in this paper. We use the conventional nomenclature ``visible'' to describe 
effectively the {\it total} matter associated with the disk and 
write the total visible matter density, $\rho_v$, as  
 $\rho_v=\rho_s+\rho_d$, the corresponding potential 
being $\Phi_v=\Phi_s+\Phi_d$. The expressions for the potentials $\Phi_s$ 
and $\Phi_d$ corresponding to the chosen forms of $\rho_s$ and $\rho_d$ 
are given in Refs. \cite{bt,co,kg}.    

Now, for the DM component, the exercise is to calculate the distribution 
of the DM particles by self-consistently including the effects of the 
self-gravitation of the DM particles themselves {\it and} the potential 
due to the total visible component specified above. The procedure we 
follow is analogous to the one developed earlier \cite{cg} with this 
difference 
that we now have to contend with the axial symmetry of the potentials. 
Since the DM particles obey the steady-state
collisionless Boltzmann equation, the assumption of Maxwellian phase-space 
density allows us to write the spatial density, $\rho_{\dm} (R,z)$, of DM as

\begin{equation}
\rho_{\dm}(R,z) =
\rho_{\dm}(0,0) \exp\biggl[-{3\over\rmsveldmsq} 
\biggl\{\biggl(\Phi_{\dm}(R,z) - 
\Phi_{\dm}(0,0)\biggr)+
\biggl(\Phi_v(R,z)-\Phi_v(0,0)
\biggr)\biggr\} \biggr],
\label{eqn-rho}
\end{equation}

\noi where the DM potential, $\Phi_{\dm}(R,z)$, satisfies the Poisson 
equation, 
\begin{equation}
\bigtriangledown^2 \Phi_{\dm}(R,z) \, = \, 4\pi{\rm G}\rho_{\dm}(R,z).
\label{eqn-poisson}
\end{equation}

\noi The solution of the coupled equations (\ref{eqn-rho})
and (\ref{eqn-poisson})  
for $\Phi_{\dm}$ is effected through the iterative scheme
($n=1,2,3,\ldots$)

\begin{equation}
\bigtriangledown^2 \phi_n (R,z)= 4\pi{\rm G}\rho_{n-1}
(R,z),
\label{eqn-discrete}
\end{equation}

\noi where $\rho_{n-1}(R,z)$ is equal to the r.h.s. of
Eq.(\ref{eqn-rho}) with 
$\Phi_{\dm}$ replaced by $\phi_{n-1}(R,z)$, and 
$\{\phi_0(R,z)-\phi_0(0,0)\} = 0$ is the initial choice for the 
iteration process. The quantities 
$\rho_{\dm}(0,0)$ and $\rmsveldm$ are taken as free parameters. 

Details of the iterative scheme and the numerical procedure are
described elsewhere. After a few iterations (typically, $n\le10$) 
the potentials $\phi_n$ converge towards the desired potential $\Phi_{\dm}$. 
We checked 
our numerical code against test equations whose exact solutions are known. 
We also check our numerical results for the actual
equations (\ref{eqn-rho}) and (\ref{eqn-poisson})   
against analytical results for small and large values of $R$ 
and $z$.  

Once $\Phi_{\dm}$ has been calculated, the rotation curve, $\Theta(R)$, 
is obtained through the relation
\begin{equation}
\Theta^2(R) \,\, = \,\, \left(R{\partial\over\partial R} 
\left[\Phi_{\dm}(R,z) + \Phi_v (R,z)\right]\right)_{z=0}.
\label{eqn-rot}
\end{equation}
\noi Note that the contribution of the visible disk to $\Theta^2(R)$ is 
proportional to its surface density [see Eq.(4-159) of Ref.\cite{bt}], 
while that of a perfect isothermal sphere is proportional to the 
square of the velocity dispersion of its 
particles [see Eq.(4-127b) of Ref.\cite{bt}].     

The theoretical rotation curves thus obtained for various
values of the parameters $\rho_{\dm}(0,0)$ and $\rmsveldm$ 
are to be compared with observations \cite{bg,fbs,ft}
to ascertain the domain of the parameter space which is acceptable.
This comparison is shown in Fig.\ref{fig} for $\Sigma_0=80\msunppcsq$ and 
$\rho_{\dm}(0,0)=1\gevpcc$. The value of $80\msunppcsq$ for $\Sigma_0$, 
it being the upper 
limit on the allowed value of $\Sigma_0$ in our calculation, gives us a 
conservative estimate of (i.e., a {\it lower limit} on) $\rmsveldm$. 
This is because, for a given value of $\Theta$ at a given value of $R$, 
a lower value of the disk surface mass density ($\Sigma_0$) requires a higher 
value of $\rmsveldm$ (for a fixed value of $\rho_{\dm}(0,0)$). 
Our choice of $\rho_{\dm}(0,0)\approx1\gevpcc$ is dictated by the 
constraint \cite{jho,jnb} that $\rho_{\dm}(R_0,0)\sim0.3\gevpcc$ and the 
need to fit 
the rotation curve. A slightly lower value of 
$\rho_{\dm}(0,0)$ generally requires higher values of $\rmsveldm$ in order 
to satisfy the above constraint and to fit  
the rotation curve. In this sense, our choice of 
$\rho_{\dm}(0,0)\approx1\gevpcc$ yields, again, a lower limit to $\rmsveldm$. 
A higher value of $\rho_{\dm}(0,0)$, on the other hand, 
can be consistent with the constraint $\rho_{\dm}(R_0,0)\sim0.3\gevpcc$ 
for sufficiently low values of $\rmsveldm$; however, in this case, the 
rotation curve falls steeply beyond the solar circle and thus provides 
a poor fit to the rotation curve.  

In order to determine (a lower limit to) the best-fit value of 
$\rmsveldm$ we have calculated 
$\chi^2\equiv{1\over N}\sum\limits_{i=1}^N\left({\Theta_i(R_i)
-\Theta_{i,o}(R_i)\over\sigma_i}\right)^2$ as a function of $\rmsveldm$  
(for $\Sigma_0=80\,\, {\rm and}\,\, 40\msunppcsq$ and $\rho_{\dm}(0,0)=
1\gevpcc$), where $N$ is the number of 
observational data points, $\Theta_i(R_i)$ and $\Theta_{i,o}(R_i)$ 
are the theoretical and observational value of the rotation speed, 
respectively, for the $i$th data point for which $R=R_i$, and $\sigma_i$ 
is the $1\sigma$ uncertainty in the measured value of 
$\Theta_{i,o}(R_i)$. We calculate the above $\chi^2$ for the entire data 
set for $R$ in the range $\sim (2$--20) kpc as well as for the restricted 
data set for $R$ in the range $\sim (10$--20) kpc in which the observed 
rotation curve data show a conspicuous rising trend. For $\Sigma_0=
80\msunppcsq$, both data set give a minimum $\chi^2$ at $\rmsveldm\sim
600\kmps$. For $\Sigma_0=40\msunppcsq$, the minimum of the $\chi^2$ lies 
at $\rmsveldm\sim 750\kmps$ for the restricted data set while the 
minimum is beyond $900\kmps$ for the full data set. From the above 
analysis we conclude that the lower limit on $\rmsveldm$ is $\sim 600\kmps$.

Notice from Fig.\ref{fig} that for $\rmsveldm\sim300\kmps$ 
the potential of the visible component concentrates
the distribution of DM towards the centre, causing the rotation
curve to fall below the observational data at large galactocentric
distances. As the kinetic energy of the DM particles increases
with increased value of $\rmsveldm$ the particles are affected
progressively less by the potentials and spread out farther. This causes
the rotation curves to be elevated.  

We thus see that the rms velocity of particles of DM needed to
generate the observed rotation curve is higher than that adopted in a
variety of discussions of DM \cite{note} . Indeed, we had an inkling that
this might be so, based on our analytic estimates made earlier in this
context \cite{rc}. The implications of this result are multifarious:

\begin{itemize}
\item{1.} Since the typical velocity of individual DM particles is higher
by at least a factor of $\sim2$ on the average, the energies they would 
deposit in
the detectors would be higher by at least a factor $\sim4$. This would make 
these events stand out against the background.

\item {2.} The higher velocities imply higher fluxes and the event rates
would be increased by at least a factor of $\sim 2$.

\item {3.} When the observed pulse height spectrum in the detectors are
reanalysed taking the above two points into account the existing bounds on
the masses and other properties of dark matter particles would become
substantially more stringent.

\item {4.} The higher velocities would also mean lower rates of capture by
the Sun by accretion; consequently the flux of high energy neutrinos
arising from their annihilations in the central regions of the
Sun \cite{steig} is expected to be correspondingly smaller.

\item {5.} The large velocities would also imply an extended halo (with an
estimated mass of $\sim 1.5\times 10^{12}\msun$ up to $\sim 100\kpc$)
whose influence on the dynamical motions within our Galaxy and on the
Local Group, as also the tidal effects on the dwarf-spheroidals would
become important. For example, this high temperature halo will impart
stability to the disk according to the criterion derived by Peebles and
Ostriker \cite{po}. 

\end{itemize}

\noi
These issues are under study and will be reported elsewhere.

\begin{figure}
\caption{
 The theoretically calculated rotation curve of the 
Galaxy for various values of $\rmsveldm$ compared with the available
observational data \protect{\cite{bg,fbs,ft}}. All curves are for 
$\rho_{\dm}(0,0)=1\gevpcc$ and 
$\Sigma_0=80\msunppcsq$ (see text). The data and error bars for $R$ in the 
range $\sim (2$--17)$\kpc$ are from Fig.3  of
Ref.\protect{\cite{fbs}}, and those for 
$R>17\kpc$ are from Fig. 2 of Ref. \protect{\cite{ft}}. The data for $R$ 
below $\sim2\kpc$ are from Ref. \protect{\cite{bg}}.  
}
\label{fig}
\end{figure}


\begin{references}
\bibitem{note} As per the suggestions received from the referee
we have repeated the calculations  with the King's model \cite{bt}
which ensures that the density of dark matter vanishes at a
finite distance. The bound on $\rmsveldm$ quoted in this paper
remain unaffected  by this change.A manuscript incorporating the
density contours ,mass distribution etc is under preparation.
\end{references}
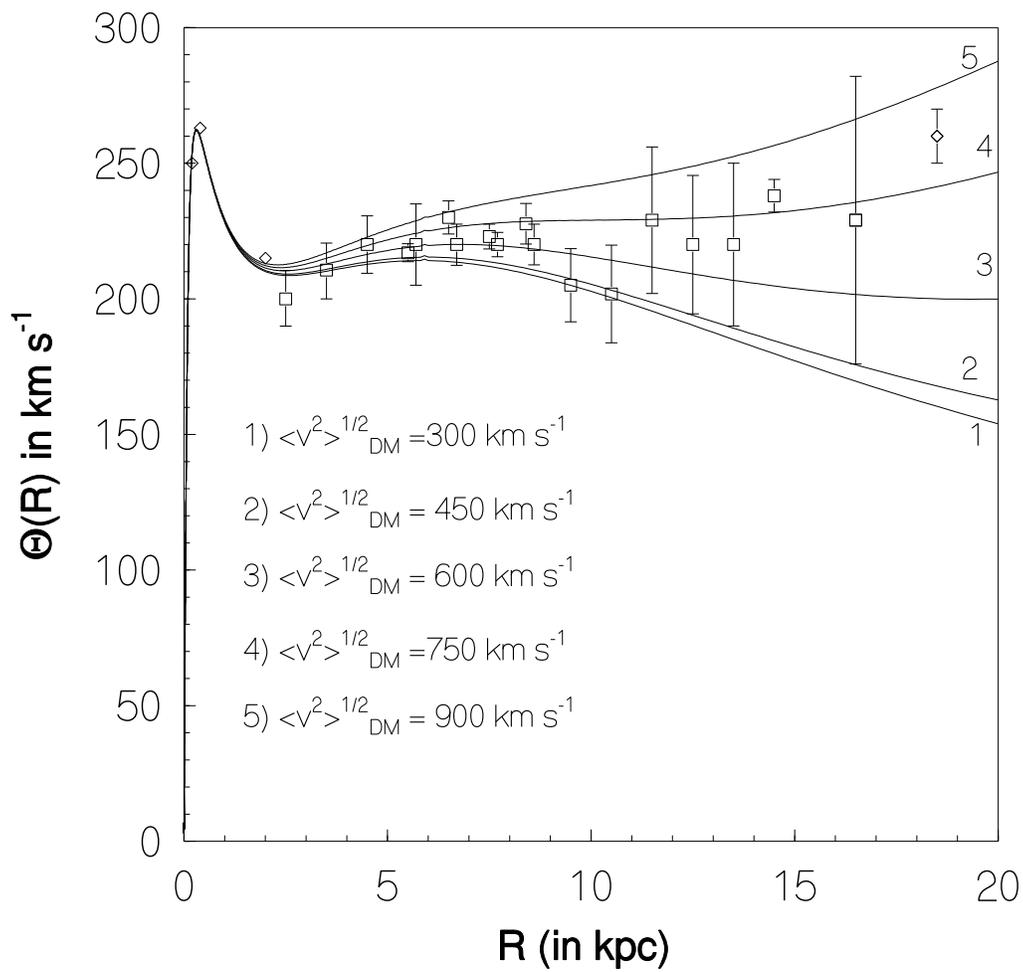
\end{document}